# Non-abelian thermal gauge potentials for high spin cold atom gases

Zheng-Chuan Wang

The University of Chinese Academy of Sciences, P. O. Box 4588, Beijing 100049, China, wangzc@ucas.ac.cn

## Abstract

On the basis of the non-equilibrium Green function formalism, we derived a spinor Boltzmann equation for the Bose cold atom gases with high spin, which is achieved by a quantum Wigner transformation on the equation satisfied by the lesser Green function. After a Taylor series expansion on the scattering terms, a temperature-dependent spinor damping force can be obtained, which can be related to a non-abelian thermal gauge potential. For the spin-1 Bose gas, the thermal gauge potential constitutes a SU(3) Lie algebra. As an example, we calculate the spin coherence oscillation for the spin-1 Bose cold atom gas trapped in the optical lattice. The relative populations in the Zeeman states as well as the temperature-dependent damping force are illustrated numerically.

# I. Introduction

In recent years, the spinor Boltzmann equation (SBE) has been a powerful tool to investigate the spin-polarized transport in spintronics[1-3], especially in the magnetic multilayers[4]. SBE was first proposed by Silin in 1957[5] based on the quantum Liouville equation, and then it was reformulated by Levy et al. to explore the spin accumulation, spin current, and spin transfer torque in magnetic multilayers[1,4]. Sheng et al. also derived a similar equation at the steady state by non-equilibrium Green function theory[2,3]. The SBE was generalized by Wang et al. to the case beyond the gradient approximation[6]; they also derived the thermal spin transfer torque and thermal spin-orbit torque under the assumption of local equilibrium[7,8] by SBE. However, the above SBEs are only suitable for describing the spin-1/2 Fermion, in particular the conduction electrons in solid, because its spinor distribution function is a $2 \times 2$ matrix, it cannot be used to study the particles with high spin, such as the ultracold quantum Fermion gases with spin $s > \frac{1}{2}$, in which the spinor distribution function should be expressed by the high-dimensional matrix.

As we know, cold atom quantum gas has achieved great progress since the realization of Bose-Einstein condensation by laser cooling in alkali atomic gases[9]. Due to its high purity and excellent control of the interaction strength, it is an ideal candidate for studying some fundamental problems in condensed matter physics, such as the superfluid, superconductor, Mott insulator, Stoner model, collective excitation et al.[10-12]. Since 2004, spinor quantum gases have attracted more and more attention, and people have observed the texture and spin dynamics in spinor Bose-Einstein condensed whatever in theory or experiment. Spinor fermion gases have also been widely explored, in particular in the spin mixing and giant spin oscillation of high-spin (s>1/2) gases[13,14].

Does the spin dynamics in high-spin Fermion or Bose gases can be described by SBE? In 2014, Sengstock et al. investigated the giant spin oscillation by SBE[14], and the relaxation dynamics of a trapped Fermion quantum gas with high spin and spatial freedom were addressed. With the exception of the relaxation procedure, the SBE could also be used to explore the spin current in Fermion gases. However, this equation is not suitable for describing the spin dynamics of Bose gases with high spin. In this manuscript, we will derive an SBE for the Bose gases with high spin and then explore its spin dynamics.

As shown by Wang[15], a temperature-dependent damping force can be derived by a Taylor series expansion on the scattering term of the quantum Boltzmann equation, which may be related to a U(1) thermal gauge potential similar to Luttinger[16] or Tatara's[17]. If we include the spin freedom in the quantum Boltzmann equation, we can obtain an SBE to describe the spin-dependent transport. For a spin of 1/2 electron, we can derive a temperature-dependent spinor damping force, which can be related to the SU(2) thermal gauge potential[18], and be expressed by the SU(2) generator $\sigma^a$ ($a = x, y, z$). For the particle with high spin, we expect that the scattering term will also contribute a temperature-dependent spinor damping force, which yields the thermal gauge potential related to other Lie algebras, i.e., for spin-1 Bose gas, the thermal gauge potential constitutes a SU(3) Lie algebra.

## II. Theoretical formalism

Let us study the cold atom gases with high spin ($s > \frac{1}{2}$), which can be described by a spinor field operator $\hat{\psi}_i(x)$ ($i = -s \ldots s$), where $i$ denotes the magnetic quantum number. To explore the spin dynamics, we should derive the SBE for the particles with high spin. As we know, the quantum Wigner distribution function satisfied by the SBE is the quantum Wigner transformation on the lesser Green function, which is

$$G_{ij}^<(x_1, x_2) = i < \hat{\psi}_i^\dagger(x_2) \psi_j(x_1) > . \quad (1)$$

If the cold atom gases are trapped in an inhomogeneous harmonic potential $V^{trap}(x) = \frac{m}{2}(\omega_x^2 x^2 + \omega_y^2 y^2 + \omega_z^2 z^2)$, an external homogeneous magnetic field is applied along the spin quantization axis, which will induce a non-linear Zeeman splitting $(qS_z^2)_{ij}$, then the second quantization Hamiltonian for this system can be written as[14]

$$H = \int dx \sum_{ij} \hat{\psi}_i^\dagger(x) [-\frac{\hbar^2}{2m} \nabla^2 \delta_{ij} + V^{trap}(x) \delta_{ij} + (qS_z^2)_{ij}] \hat{\psi}_j(x) +$$

$$\frac{1}{2} \sum_{ijkl} U_{ijkl} \hat{\psi}_i^\dagger(x) \hat{\psi}_k^\dagger(x) \hat{\psi}_l(x) \hat{\psi}_j(x), \quad (2)$$

where the second term accounts for the interaction part of the scattering potentials, for the even total spin F, the coupling constant $U_{ijkl} = \sum_{S=0}^{2F-1} g_S \sum_{M=-S}^{S} <ik|SM><SM|jl>$, $g_S = \frac{4\pi\hbar^2 a_S}{m}$ is the scattering strength, $a_S$ denotes the scattering length, $<SM|jl>$ represent the Clebsh-Gordon matrix elements. By means of the Heisenberg equation $i\hbar \frac{\partial \hat{\psi}_m(y)}{\partial t} = [\hat{\psi}_m(y), H]$ and the commutation relation $[\hat{\psi}_i^\dagger(x), \hat{\psi}_j(y)] = \delta_{ij}\delta(x-y)$ or the anti-commutation relation $\{\hat{\psi}_i^\dagger(x), \hat{\psi}_j(y)\} = \delta_{ij}\delta(x-y)$ for the Bose or Fermion, respectively, we can accomplish the equation obeyed by the field operator as

$$i\hbar \frac{\partial \hat{\psi}_m(y)}{\partial t} = -(-\frac{\hbar^2}{2m}\nabla^2 + V^{trap}(x))\hat{\psi}_m(y) -$$

$$q \sum_j (S_z)_{mj}^2 \hat{\psi}_j(y) -$$

$$\sum_{ijl} U_{ijml}(\hat{\psi}_i^\dagger(y)\hat{\psi}_l(y)\hat{\psi}_j(y)), \quad (3)$$

By use of Eq.(3), the motion equation for the lesser Green function can be written as

$$\hbar \frac{\partial G_{ij}^<(x_1, x_2)}{\partial t_1} = -(-\frac{\hbar^2}{2m}\nabla_{x_1}^2 + V^{trap}(x_1))G_{ij}^<(x_1, x_2) -$$

$$q \sum_k (S_z)_{jk}^2 G_{ik}^<(x_1, x_2) -$$

$$\sum_{mnl} U_{mnjl}(G_{in}^<(x_1, x_2)G_{ml}^<(x_1, x_1) \pm$$

$G_{il}^<(x_1,x_2)G_{mn}^<(x_1,x_1))$, (4)

where we have adopted the Hartree-Fock approximation as

$<\hat{\psi}_n^\dagger(x_2)\hat{\psi}_l^\dagger(x_2)\hat{\psi}_m(x_2)\hat{\psi}_j(x_1)>\approx<\hat{\psi}_n^\dagger(x_2)\hat{\psi}_j(x_1)><\hat{\psi}_l^\dagger(x_2)\hat{\psi}_m(x_2)>\pm<\hat{\psi}_n^\dagger(x_2)\hat{\psi}_m(x_2)><\hat{\psi}_l^\dagger(x_2)\hat{\psi}_j(x_1)>$, (5)

in which "+" corresponds to the Bose, while "−" corresponds to the Fermion. Similarly, by use of the conjugate of Eq.(3), we have

$\hbar\frac{\partial G_{ij}^<(x_1,x_2)}{\partial t_2}=\left(-\frac{\hbar^2}{2m}\nabla_{x_2}^2+V^{trap}(x_2)\right)G_{ij}^<(x_1,x_2)+q\sum_k(S_z)_{ik}^2 G_{kj}^<(x_1,x_2)+\sum_{mnl}U_{mnil}(G_{nj}^<(x_1,x_2)G_{ml}^<(x_2,x_2)\pm G_{nm}^<(x_2,x_2)G_{lj}^<(x_2,x_1))$, (6)

Subtract Eq.(3) by its conjugate form Eq.(6), and use the center coordinates and relative coordinates $r=x_1-x_2$, $R=x_1+x_2$, $t=t_1-t_2$, $T=t_1+t_2$, we have

$\hbar\frac{\partial G_{ij}^<(R+\frac{r}{2},R-\frac{r}{2})}{\partial T}=-(-\frac{1}{m}\nabla_r\nabla_R+\frac{1}{2}m\omega^2 rR)G_{ij}^<(R+\frac{r}{2},R-\frac{r}{2})-q\sum_k(S_z)_{jk}^2 G_{ik}^<(R+\frac{r}{2},R-\frac{r}{2})+q\sum_k(S_z)_{ik}^2 G_{kj}^<(R+\frac{r}{2},R-\frac{r}{2})-\sum_{mnl}U_{mnjl}(G_{in}^<\left(R+\frac{r}{2},R-\frac{r}{2}\right)G_{ml}^<\left(R+\frac{r}{2},R+\frac{r}{2}\right)\pm G_{il}^<\left(R+\frac{r}{2},R-\frac{r}{2}\right)G_{mn}^<\left(R+\frac{r}{2},R+\frac{r}{2}\right))-\sum_{mnl}U_{mnil}(G_{nj}^<(R+\frac{r}{2},R-\frac{r}{2})G_{lm}^<(R-\frac{r}{2},R-\frac{r}{2})\pm G_{nm}^<(R-\frac{r}{2},R-\frac{r}{2})G_{lj}^<(R+\frac{r}{2},R-\frac{r}{2}))$,(7)

As we know, the spinor distribution function is defined as the quantum Wigner transformation on the lesser Green function as follows:

$f_{il}(p,R,T)=\frac{i}{2\pi\hbar}\int dr\, exp(-\frac{ip}{\hbar}r)G_{il}^<(R+\frac{r}{2},R-\frac{r}{2})$, (8)

if we make a quantum Wigner transformation on both sides of Eq.(7) over the relative coordinate, we have

$\frac{\partial f_{ij}(p,R,T)}{\partial T}=\left(-v\nabla_R+\frac{1}{2}m\omega^2 R\frac{\partial}{\partial p}\right)f_{ij}(p,R,T)-q\sum_k(S_z)_{jk}^2 f_{ik}(p,R,T)+q\sum_k(S_z)_{ik}^2 f_{kj}(p,R,T)-\sum_{mnl}U_{mnjl}(\int dj\, J_{1ml}(j,R)f_{in}(p+j,R,T)\pm\int dj\, J_{1mn}(j,R)f_{il}(p+j,R,T)-\sum_{mnl}U_{mnil}\int dj\, J_{2lm}(j,R)f_{nj}(p+j,R,T)\pm\int dj\, J_{2nm}(j,R)f_{lj}(p+j,R,T))$, (9)

where $J_{1mn}(j,R)=\int dz\, exp(\frac{ij}{\hbar}z)G_{mn}^<(R+\frac{z}{2},R+\frac{z}{2})$ and $J_{2lm}(j,R)=\int dz\, exp(\frac{ij}{\hbar}z)G_{lm}^<(R-\frac{z}{2},R-\frac{z}{2})$ can be interpreted as the probability of a jump in the momenta with the amount $j$, where we have used the Wigner formula given in Ref.[19] to handle the convolution on the quantum Wigner transformation. In Eq.(9), "−" corresponds to the Fermion, then the last four terms that describe the scattering of cold atoms will cancel with each other, which is similar to the SBE given by Sengstock[14]. In Sengastock's paper, they derived a collisionless Boltzmann equation in the Hartree-Fock approximation. In our method, the scattering terms also become to zero in the Hartree-Fock approximation. When we choose the sign "+" in Eq.(9), it corresponds to the Bose, in this case the scattering terms will exist, so our SBE (9) is only suitable to describe the Bose, which is the starting point for our next work.

### III. The temperature dependent thermal gauge potential

Next, we try to derive the temperature-dependent damping force for the Bose as Ref.[15]. By adopting the Taylor series expansion on the spinor distribution function in the scattering term, i.e. $f_{il}(p+j,R,T)$ in the integral $I = \int dj\, J_{1mn}(j,R) f_{il}(p+j,R,T)$, we have

$$f_{il}(p+j,R,T) = f_{il}(p,R,T) + \frac{\partial f_{il}}{\partial p} j + \frac{1}{2}\frac{\partial^2 f_{il}}{\partial p^2} j^2 + \cdots, \quad (10)$$

then the integral $I = \int dj\, J_{1mn}(j,R) f_{il}(p+j,R,T)$ can be expressed as

$$I = f_{il}(p,R,T) \int dj\, J_{1mn}(j,R) + \frac{\partial f_{il}}{\partial p} \int dj\, j J_{1mn}(j,R) + \frac{1}{2}\frac{\partial^2 f_{il}}{\partial p^2} \int dj\, j^2 J_{1mn}(j,R) + \cdots. \quad (11)$$

If we only keep the first-order term and neglect the higher-order terms in the above integral, Eq.(9) for the Bose can be written as

$$\frac{\partial f_{ij}(p,R,T)}{\partial T} = \left(-v\nabla_R + \frac{1}{2} m\omega^2 R \frac{\partial}{\partial p}\right) f_{ij}(p,R,T) - q\sum_k (S_z)^2_{jk} f_{ik}(p,R,T) + q\sum_k (S_z)^2_{ik} f_{kj}(p,R,T) - 2\sum_{mnl} U_{mnjl} \left(\int dj\, j J_{1mn}(j,R) \frac{\partial f_{il}}{\partial p}\right) - 2\sum_{mnl} U_{mnil} \left(\int dj\, j J_{2lm}(j,R) \frac{\partial f_{nj}}{\partial p}\right) - 2\sum_{mnl} U_{mnjl} \left(\int dj\, J_{1mn}(j,R) f_{il}\right) - 2\sum_{mnl} U_{mnil} \left(\int dj\, J_{2lm}(j,R) f_{nj}\right), \quad (12)$$

The total spin s during the scattering of two atoms is conserved, $i+j = k+l = s$. If we define the element $F_{in}$ of matrix $\hat{F}$ as

$$F_{in} = 2\sum_{ml} U_{mnil} \int dj\, j J_{2lm}(j,R), \quad (13)$$

and the element $\tau^{-1}{}_{in}$ of matrix $\hat{\tau}^{-1}$ as

$$\tau^{-1}{}_{in} = 2\sum_{ml} U_{mnil} \int dj\, J_{2lm}(j,R), \quad (14)$$

Eq.(12) can be further expressed in the matrix form:

$$\frac{\partial \hat{f}(p,R,T)}{\partial T} + \left(v\nabla_R - \frac{1}{2} m\omega^2 R \frac{\partial}{\partial p}\right) \hat{f}(p,R,T) - q[S_z^2, \hat{f}] + \hat{F}\frac{\partial \hat{f}}{\partial p} + \frac{\partial \hat{f}}{\partial p}\hat{F}^\dagger = -\hat{\tau}^{-1}\hat{f} - \hat{f}\hat{\tau}^{-1\dagger}. \quad (15)$$

Similar to Ref.[15,18], the coefficient $\hat{F}$ is named as the spinor damping force. $\hat{\tau}^{-1}$ is named as the inverse relaxation time, which is a generalization of the usual relaxation time constant. For spin-1 Bose, $\hat{F}$ and $\hat{\tau}^{-1}$ are $3\times 3$ matrix, so we can expand them by the complete matrix basis formed by the unit matrix $\hat{I}$ and 8 SU(3) generators $\hat{T}_i$ ($i=1\ldots 8$), i.e. for the damping force $\hat{F}$, we have

$$\hat{F} = \vec{F}_0 \hat{I} + \sum_i \vec{F}_i \hat{T}_i. \tag{16}$$

Since the lesser Green function in $J_{2lm}(j, R)$ of Eq.(13) can be related to the density of the particle, after adopting the local equilibrium assumption, we can expand the lesser Green function around the local equilibrium distribution $f^0(p, R)$ as

$$-i\hbar G_{ml}^<(R, R) = \int dp [f^0(p, R) +$$

$$(\frac{\partial f^0}{\partial \varepsilon}) f_{ml}(P, R, T) + \ldots], \tag{17}$$

where $f_{ml}(P, R, T)$ is the first-order distribution function deviating from equilibrium. Neglecting the higher-order terms, the damping force in Eq.(13) can be expressed as

$$F_{in} =$$
$$-\frac{2}{i\hbar} \{ \int dp [\sum_{ml} U_{mnil} \int dj\, j \int dz\, exp(\frac{ij}{\hbar}z)(f^0(p, R - \frac{z}{2}) + (\frac{\partial f^0}{\partial \varepsilon}) f_{ml}(P, R - \frac{z}{2}, T))] \}.$$

$$\tag{18}$$

Since the equilibrium distribution function $f^0(p, R)$ contain a temperature $T(x)$, then the spinor damping forces are temperature dependent.

The temperature-dependent damping force can be related to the thermal scalar and vector potentials as Ref.[15,18]. The force $\vec{F}_0$ in the spinor damping force Eq.(16) have its non-zero divergence $\vec{\nabla} \cdot \vec{F}_0$ and curl $\vec{\nabla} \times \vec{F}_0$, so it is a dissipative force. If we introduce the thermal scalar potential $\varphi$ as

$$\nabla^2 \varphi = \vec{\nabla} \cdot \vec{F}_0, \tag{19}$$

and the vector potential as

$$-\frac{\partial (\vec{\nabla} \times \vec{A})}{\partial t} = \vec{\nabla} \times \vec{F}_0. \tag{20}$$

We can see that the thermal scalar and vector potentials can be related with the equilibrium distribution function $f^0(p, R)$ and the distribution function $f_{ml}(P, R, T)$ according to Eq.(18), which implies that they are also temperature dependent.

If we make a U(1) gauge transformation $\varphi \to \varphi - \dot{\chi}$ and $\vec{A} \to \vec{A} + \nabla \chi$ on Eq.(19) and (20), where $\chi$ is a scalar function, they can keep the gauge invariant, and there exists gauge freedom for the thermal potentials $\varphi$ and $\vec{A}$. Therefore the damping force $\vec{F}_0$ is gauge invariant. Our damping force originates from the interaction in the scattering term of the cold atoms, and it is not the usual U(1) gauge potential caused by the external electric field.

As we know, $\hat{T}_i$ ($i = 1\ldots 8$) is the generator of SU(3) group, so $\vec{F}_i$ in the spinor damping force Eq.(16) can be related to the SU(3) gauge potential $\vec{A} = A^i \hat{T}_i (i = 1\ldots 8)$. If we introduce the gauge strength tensor as $F_{\mu\nu}^a = \partial_\mu A_\nu^a - \partial_\nu A_\mu^a + e\epsilon_{abc} A_\mu^b A_\nu^c$, where $e$ is the interaction constant, and $\epsilon_{abc}$ is the Kronecker symbol. The SU(3) gauge potential has the following Lagrangian

$$\mathcal{L}_{su(3)} = -\frac{1}{4}\text{Tr}(F_{\mu\nu}F_{\mu\nu}), \quad (21)$$

which should be equal to $\mathcal{L}_{su(3)} = \frac{\vec{p}^2}{2m}\hat{I} - \hat{V}$, where

$$\hat{V} = e\int \sum_i \vec{F}_i \hat{T}_i \cdot d\vec{x}, \quad (22)$$

then we can relate the $\vec{F}_i$ in the spinor damping force with the SU(3) gauge potential as

$$\frac{\vec{p}^2}{2m}\hat{I} - e\int \sum_i \vec{F}_i \hat{T}_i \cdot d\vec{x} = -\frac{1}{4}\text{Tr}(F_{\mu\nu}F_{\mu\nu}). \quad (23)$$

Therefore the above SU(3) gauge potential $\vec{A} = A^i \hat{T}_i$ can be related with the equilibrium distribution function $f^0(p, R)$ and the distribution function $f_{ml}(P, R, T)$, it is also temperature dependent.

There exists a gauge freedom for the above equation. The damping force maintains the gauge invariance under the SU(3) gauge transformation. On the other hand, our spinor damping force originates from the interaction in the scattering terms of the cold atoms, which is different from the real SU(3) strong interaction in particle physics.

## IV. Numerical Results

It is difficult to find the analytical solutions for Eq.(12) due to its position dependent coefficients. Eq.(12) is a differential and integral equation group satisfied by the spinor distribution function. If we adopt the single-mode approximation as Ref.[14], we have

$$f_{ij}(p, R, T) = f(p, R)\rho_{ij}(T), \quad (24)$$

where $\rho_{ij}(T)$ describes the spin configuration of the spinor distribution in the homogeneous phase space, then Eq.(12) become to

$$\frac{\partial \rho_{ij}(T)}{\partial T} = \frac{1}{f(p,R)}\{(-v\nabla_R + \frac{1}{2}m\omega^2 R\frac{\partial}{\partial p})f(p,R) -$$

$$qf(p,R)\sum_k (S_z)^2_{jk}\rho_{ik}(T) +$$

$$qf(p,R)\sum_k (S_z)^2_{ik}\rho_{kj}(T) -$$

$$2\sum_{mnl} U_{mnjl}\left(\int djj J_{1mn}(j,R)\rho_{il}(T)\frac{\partial f}{\partial p}\right) -$$

$$2\sum_{mnl} U_{mnil}\left(\int dj j J_{2lm}(j,R)\rho_{nj}(T)\frac{\partial f}{\partial p}\right) -$$

$$2f(p,R)\sum_{mnl} U_{mnjl}\left(\int dj J_{1mn}(j,R)\rho_{il}(T)\right) -$$

$$2f(p,R)\sum_{mnl} U_{mnil}\left(\int dj J_{2lm}(j,R)\rho_{nj}(T)\right),$$

$$(25)$$

If we simply approximate $f(p, R)$ on the right hand side of Eq.(25) as the local equilibrium distribution function $f^0(p, R)$, we can obtain a linear equation group for $\rho_{ij}(T)$, which can be solved after obtaining its eigenvalues and eigenvectors, then the total distribution $f_{ij}(p, R, T)$ can be obtained approximately.

As an example, we investigate the spin dynamics of $s = 1$ $^{87}$Rb spinor Bose gases trapped in a magnetic harmonic potential with the frequency $2\pi \times 15 Hz$. For simplicity, we only study the one-dimensional case. In our calculation, an external magnetic field B $=0.28$G is applied along the spin quantization axis[20], which produces a non-zero Zeeman splitting with $2q \propto B^2$.

The relative populations in the Zeeman states are shown in Fig. 1 and Fig. 2. In Fig. 1, we demonstrate the relative population in $|s, m_s; s,$

$m_s>=|1,0;1,0>$ Zeeman state, where $m_s$ is the magnetic quantum number. Fig. 2 shows the relative population in the $|s, m_s ;s, m_s>=|1,+1;1-1>$ Zeeman state. Both of them oscillate with time at a frequency of 40.2 Hz, which are just the spin coherence oscillations. The curve of the line-symbol given the experimental data from Ref.[20], the solid line is the theoretical result given by us, which is conceded to the experimental data. The oscillations of the population in $|1,0;1,0>$ and $|1,+1;1-1>$ Zeeman states have a phase difference π, it indicates that the spin coherence oscillation is in essence a Rabi oscillation. Meanwhile, the oscillations have some decrease in amplitude, which is caused by the relaxation procedure.

The temperature-dependent damping force as a function of position is shown in Fig. 3. This force is a 3×3 matrix, which has nine components. If we adopt the single-mode approximation as Eq.(24), the nine components have a common momenta and position dependence described by the function $f(p,R)$, while the spin-dependent part are contained in the time-dependent part $\rho_{ij}(T)$. The oscillations of the damping force with time are similar to the relative populations shown in Figs. 1 and 2, so we don't exhibit it again. Here, we are only interested in the position and temperature dependence of the damping force at a certain time t = 16 ms, so all the nine components have a similar shape except the magnitude caused by the time dependence at t = 16 ms. In Fig. 3, we only plot the first component of the spinor damping force at different temperatures. We can see that the damping force increases with position, physically it originates from the spin-dependent scattering of the cold atoms. The higher of the temperature, the larger of the damping force.

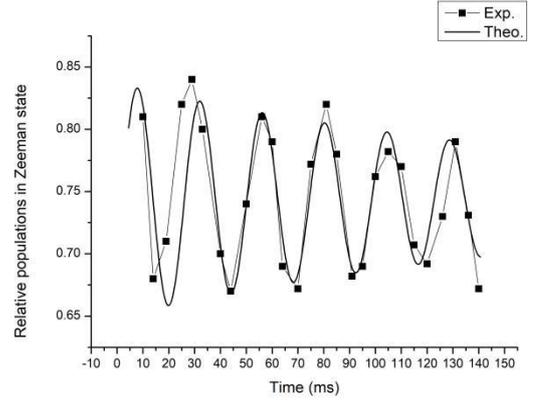

Fig1. The relative population in $|1,0; 1,0>$ Zeeman state vs time.

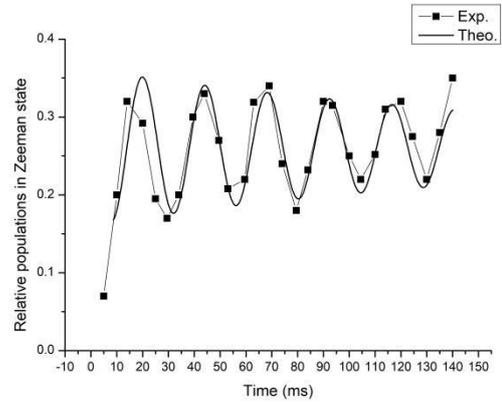

Fig.2 The relative population in $|1,+1; 1-1>$ Zeeman state vs time.

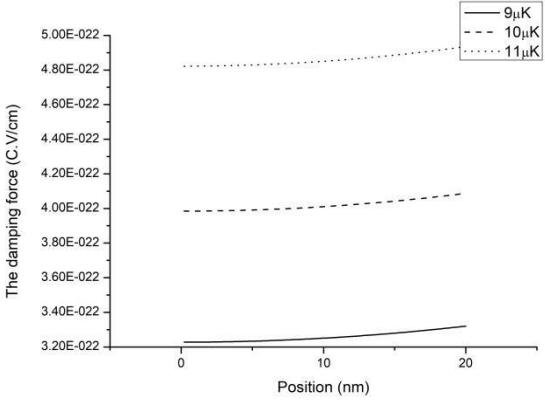

Fig.3 The damping force vs position at different temperatures T = 9μK, 10μK, 11μK.

## IV. Summary and discussions

By extending the spin-1/2 SBE to the case of high spin, we investigated the spin dynamics of the cold atoms when an external homogeneous magnetic field was applied in the optical lattice. The spin coherence oscillation and the temperature-dependent damping force are shown in Figs.1-3, respectively. The temperature-dependent spinor damping force originates from the spin-dependent scattering of the cold atoms, so it is in essence an electromagnetic force. Since it is a dissipative force with the curl and divergence, we relate its $\vec{F}_0$ part with the U(1) scalar and vector potential, and the $\vec{F}_i$ ($i$=1...8) part with the SU(3) gauge potential for spin-1 Bose. All of the gauge potentials are temperature dependent; therefore, they are thermal gauge potentials.

Although the thermal gauge potential for spin-1 Bose is a SU(3) gauge potential, it is different from the SU(3) gauge potential in quantum chromodynamics for the strong interaction of Quarks. Our SU(3) thermal gauge potential comes from the electromagnetic force, and the gauge potential is induced by the spin-dependent scattering of the cold atoms. Our thermal gauge potential is also different from the artificial gauge potential induced in cold atom physics. The latter is caused by the rotation of the magnetic trap or by the geometric phase produced by the motion of a neutral particle with spin in the inhomogeneous magnetic field, while our non-abelian thermal gauge potential is concerned with the spin freedom.

### Acknowledgments

This study is supported by the National Key R&D Program of China (Grant No. 2022YFA1402703).